\documentclass[lettersize,journal]{IEEEtran}
\usepackage{amsmath,amsfonts}
\usepackage{amssymb}
\usepackage{algorithmic}
\usepackage{algorithm}
\usepackage{array}
\usepackage[caption=false,font=normalsize,labelfont=sf,textfont=sf]{subfig}
\usepackage{textcomp}
\usepackage{stfloats}
\usepackage{url}
\usepackage{verbatim}
\usepackage{graphicx}
\usepackage{cite}
\begin{document}

\title{A Technique for Load Shifting Low-latency Applications in Multi-Region Renewables Harvesting via SMT Core Pooling}

\author{Tharindu B. Hewage, Shashikant Ilager, Maria A. Rodriguez, and Rajkumar Buyya
\vspace{-30pt}
}

\markboth{}{}

\maketitle

\begin{abstract}
Load shifting across geographic regions to chase intermittent renewable energy availability is commonly used in reducing cloud infrastructure carbon footprint. However, it often omits low-latency applications due to high latency variances of wide area networks (WAN) that interconnect regions. This paper addresses accommodating low-latency applications into load shifting by minimizing their shifting across the WAN. We propose a technique using a hardware-software co-design approach. At the hardware level, we conduct server load matching over renewables supply peaks and valleys by deep idling physical cores in two otherwise identical server pools, with one enabling simultaneous multi-threading (SMT) in CPUs. In return, we achieve a static set of logical cores amidst energy supply dynamics, reducing the probability of workload shifting. At the software level, we efficiently chase the static set of cores for low-latency applications within regions while prioritizing best-effort applications to accommodate shifting requirements across WANs. Our approach exploits the lower performance compromise of SMT cores due to their hardware multi-threading. We implement the proposed technique with OpenStack and CPU idle states and evaluate its performance on a real experimental testbed with Azure VM arrival traces. Results show an 80\% reduction in offloading low-latency VMs and a 43.81\% reduction in coefficient of variation of p90 end-user latency while having a worst-case latency compromise of 11.97\% due to SMT cores.
\end{abstract}


\section{Introduction}

The data center industry is expected to reach nearly 13\% of global electricity demand by 2030 \cite{carbon-free-google20}. With electricity generation being one of the most significant contributions to global CO$_{2}$ emissions \cite{hawken_drawdown_reverseglobalwarming2017}, data center operators are presented with the challenge of grid decarbonization. Cloud data center fleets are often spatially distributed and powered by sources integrated with intermittent renewables, emitting varying amounts of CO$_{2}$ per kilowatt-hour (kWh) \cite{radovanovic23datacentercarbonaware}. As a result, carbon optimization in cloud computing infrastructures predominantly relies on load shifting$-$shifting workloads across data center locations via cloud Virtual Machine (VM) scheduling \cite{masanet2020recalibratingdatacenterenergyestimates, radovanovic23datacentercarbonaware, liu2011greeningloadbalance, zheng2020mitigating, james2019lowcarbonkubenetes}$-$to chase availability of renewables for workload execution. Adversely, network traffic among spatially distributed data centers can hop through Wide Area Networks (WAN) with high latency variances \cite{miao24megate}, resulting in unpredictable end-user latency performances in VMs. For instance, when aggregated network traffic of VMs reaches edge routers of WAN, traffic management at the IP layer that is unaware of the aggregation can split them into tunnels with different latency performances \cite{miao24megate}. To avoid potential application performance degradation from such unpredictable latency performances, cloud providers often limit load shifting to best-effort workloads with flexible latency requirements, such as scientific simulations, batch processing, and machine learning training \cite{radovanovic23datacentercarbonaware}. To this end, this work explores the opportunities in accommodating low-latency workloads for cloud load shifting, emphasizing minimizing the high latency variance that VMs experience during load shifting.

\begin{figure}[tpb]
    \centering
    \includegraphics[width=\linewidth]{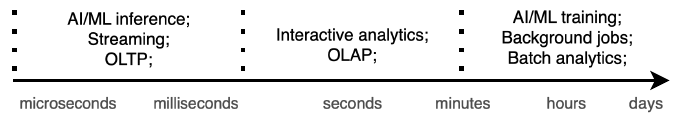}
    \caption{Low-latency applications in cloud data centers \cite{anderson23treehouse}.}
    \label{fig::latency-spectrum}
\end{figure}

Use cases of cloud low-latency computing are emerging across various industries, including healthcare, factories, automotive, and aviation \cite{ubuntu2020ctoguideforrealtimekernel}. According to recent forecasts, they are predicted to accommodate nearly 30\% of the world's data in the near term \cite{idc2018digitizationofworld}. Today, low-latency computing plays a vital role in content delivery networks, streaming applications, and applications of recent AI-boom, such as serving generative AI models. Figure \ref{fig::latency-spectrum} illustrates a spectrum of such applications available in the cloud according to their latency service level agreements (SLAs). Preserving latency SLAs is imperative to maintain application service quality, whereas accommodating the growing low-latency applications in load shifting is equally important for the sustainable growth of clouds. The key challenge is the workload impact from cloud network overheads. At renewables supply valleys, the data center's workload must be migrated to match the server's reduced resource capacity. As migrations often move workloads across WANs, workloads are then susceptible to WAN's high latency variances. Existing techniques for load shifting either compromise latency performance with potential server performance throttling \cite{jahanshahi2022powermorph} or do not consider opportunities in limiting workload migration within the data center.

To address these gaps, we propose a technique that aims to contain low-latency applications within the local cloud region. We leverage a hardware-software co-design approach. At the hardware level, we tackle the server resource capacity reduction problem in renewable supply valleys. We apply an application-independent CPU core-level power management mechanism with two heterogeneous server pools. CPUs in server pools are physically the same, yet one enables simultaneous multi-threading (SMT) to double the available server logical cores through hardware multi-threading. During supply valleys, half of the CPU cores are set to deep idle and restored at the peaks. In return, we maintain a static set of logical cores across the server pools. At supply valleys, servers of the SMT pool exhibit the static set, whereas the servers of the non-SMT pool exhibit the static set at supply peaks. At the software level, we dynamically chase the static set of logical cores across the server pools for low-latency workloads. In our technique, load shifting for low-latency workloads is mostly conducted within the local cloud region using its fast network fabric, eliminating the communication overheads of WAN. Further, executing low-latency workloads in the SMT pool incurs only a minimum performance overhead since SMT cores use hardware multi-threading, which provides better performance.

We implement our technique with OpenStack and core-level power management with CPU idle states. To evaluate, we use an experimental cloud region with SMT server pooling and a local network fabric. We use an HP ProLiant server for each pool with a 12-core Intel Xeon silver CPU. We enable SMT through Intel Hyper-threading technology. We use VM arrival data from Azure's workload traces and renewable dynamics from ELIA solar data. We measure low-latency performance inside VMs by running the Cyclictest tool. The \textbf{key contributions} of our work are as follows:

\noindent (1) Propose a new technique for localized load shifting of low-latency workloads to integrate renewable energy, avoiding the latency compromises of workload shifting over WANs across geographical regions.\\
\noindent (2) Implement the proposed technique in real cloud settings and conduct detailed experiments using production VM and renewable energy data.\\
\noindent (3) Evaluate the proposed technique against state-of-the-art baselines, focusing on maintaining low-latency performance. Our results show an 80\% reduction in offloading low-latency VMs, a 43.81\% reduction in coefficient of variation of p90 end-user latency, and an 11.97\% performance compromise due to SMT cores.

\section{Background and Motivation}
\label{sec::background-and-motivation}

In this section, we provide background on load shifting for the utilization of renewable energy across geographically distributed cloud regions, introducing its dynamic server resource scaling problem and challenges in accommodating low-latency applications. We then detail our motivations for solving those.

\begin{figure}[tpb]
    \centering
    \includegraphics[width=1.0\linewidth]{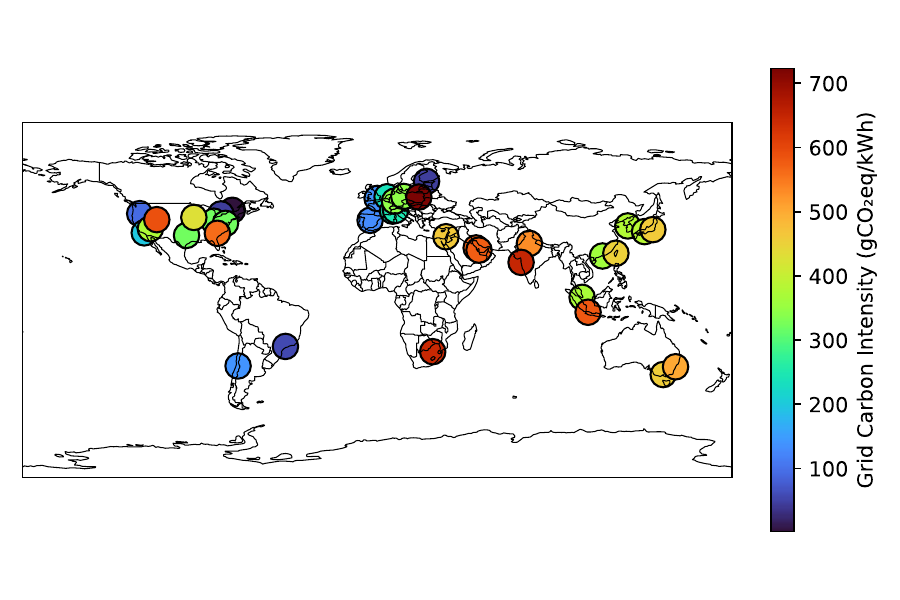}
    \caption{Average operational gross carbon emissions per unit of energy from the power grid across Google cloud regions \cite{google25clour-carbon-regions}.}
    \label{fig::mt-google-carbon}
\end{figure}

\noindent\textbf{Dynamic server resource scaling problem in utilizing intermittent renewable energy availability across cloud regions:} Cloud providers provision data centers across geographical regions around the globe to cater to various requirements, such as end-user latency and data privacy. \cite{google25clour-carbon-regions}. Those regions are powered by electricity grids with varying carbon intensities in their supply, primarily due to the integration of intermittent renewable energy sources such as solar and wind \cite{hanafy23carbonscaler}. Recently, driven by 24/7 zero emission goals \cite{carbon-free-google20}, hyper-scale cloud providers increasingly utilize low-carbon computing opportunities across their cloud regions \cite{radovanovic23datacentercarbonaware}. Figure \ref{fig::mt-google-carbon} illustrates grid carbon intensity per unit of energy across Google cloud regions. In that, executing workloads with lower carbon intensity is beneficial in reducing cloud providers' carbon footprint. In doing so, cloud platforms model carbon intensity as a resource for optimization. Since the carbon intensity of cloud regions predominately depends on the supply dynamics of renewable energy sources, it can be modeled as a dynamic server resource scaling problem, where server resource capacity scales up or down depending on the supply dynamics of clean energy. Recent examples of such can be seen in virtual capacity caurves \cite{radovanovic23datacentercarbonaware} and workload scaling for carbon-optimization \cite{hanafy23carbonscaler}. The key approach in addressing dynamic server resource scaling is \emph{load shifting}, where the execution of workloads is either shifted in time or space across cloud regions.

\begin{figure}[tpb]
    \centering
    \includegraphics[width=1.0\linewidth]{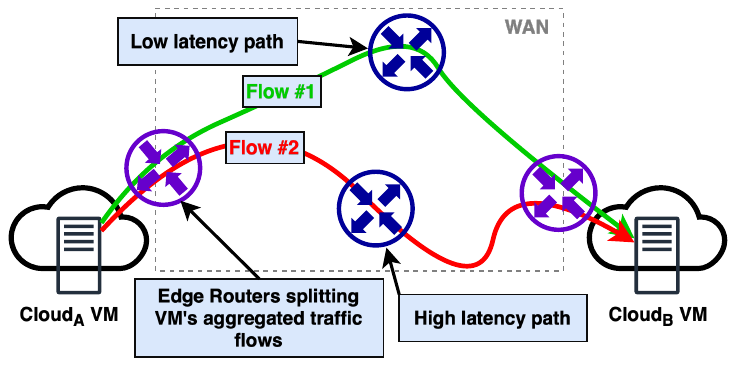}
    \caption{Traffic management at IP layer splitting aggregated traffic flows of cloud VMs when communicating over WAN \cite{miao24megate}.}
    \label{fig::mt-traffic-splitting}
\end{figure}

\noindent\textbf{Challenges in accommodating low-latency applications for load shifting:} Low-latency applications require maintaining an upper bound in their end-user communication latency. Figure \ref{fig::latency-spectrum} illustrates a spectrum of end-user latency for various low-latency applications in the cloud. Such applications are increasingly prevalent \cite{idc2018digitizationofworld}; thus, their impact on carbon-aware cloud computing is growing. However, low-latency applications are not flexible in load shifting, primarily due to the complexities in maintaining their rigid performance constraints amidst server resource down-scaling of renewable energy supply valleys. These applications cannot tolerate longer service unavailabilities in shifting across time, and more importantly, they cannot tolerate shifting across space (i.e., shifting among cloud regions) due to unpredictability in network performance connecting the cloud regions. Figure \ref{fig::mt-traffic-splitting} illustrates an example of that. Geographically distributed cloud regions are connected using wide area networks (WAN). Each cloud region connects to the WAN using an edge router and passes its network traffic flows to the WAN's traffic management. Recent studies show that WAN's traffic management at the IP layer is unable to distinguish aggregated traffic flows of a cloud virtual machine; thus, an aggregated traffic flow can get split at the edge router, routed through different paths with various latency performances before arriving at the destination \cite{miao24megate}. The resulting application end-user latency is unpredictable and unfavorable to application users \cite{kumhare2021poweroversubscription}. Further, increased latency performance may violate application service level agreements (SLAs), which can incur costly penalties to the cloud provider.

\noindent\textbf{Motivations:} Therefore, to accommodate low-latency applications in load shifting, we identify server resource down-scaling at renewable energy supply valleys as its major bottleneck. Since load shifting must offload workload outside the local cloud region to match scaled-down server resource capacity, we \textbf{\emph{hypothesize that maintaining a static resource capacity amidst renewable supply valleys for low-latency applications would enable load shifting to retain those within the local cloud regions}}, avoiding costly communication overheads of WANs. In this context, we outline the following research questions to draw our motivations for this paper.

\par\noindent\textit{\textbf{o1}: Can we manage static server resource capacity amidst renewables valleys to yield an infrastructure favorable for low-latency applications?}
\par\noindent\textit{\textbf{o2:} For a technique designed to achieve o1, what compromises are made, and more importantly, can the compromises operate within the service level agreements (SLAs) of low-latency applications?}

In addressing \textit{o1} and \textit{o2}, we propose a load-shifting technique to accommodate low-latency applications in carbon-aware cloud computing. Section \ref{sec::technique} details its inner workings, Section \ref{sec::implementation} describes our implementation of the proposed technique, and Section \ref{sec::performance-evaluation} evaluates and discusses its performance over state-of-the-art baselines.

\section{System Model and Problem Formulation}
\label{sec::system-model}

\begin{figure*}[tpb]
    \centering
    \includegraphics[width=0.9\linewidth]{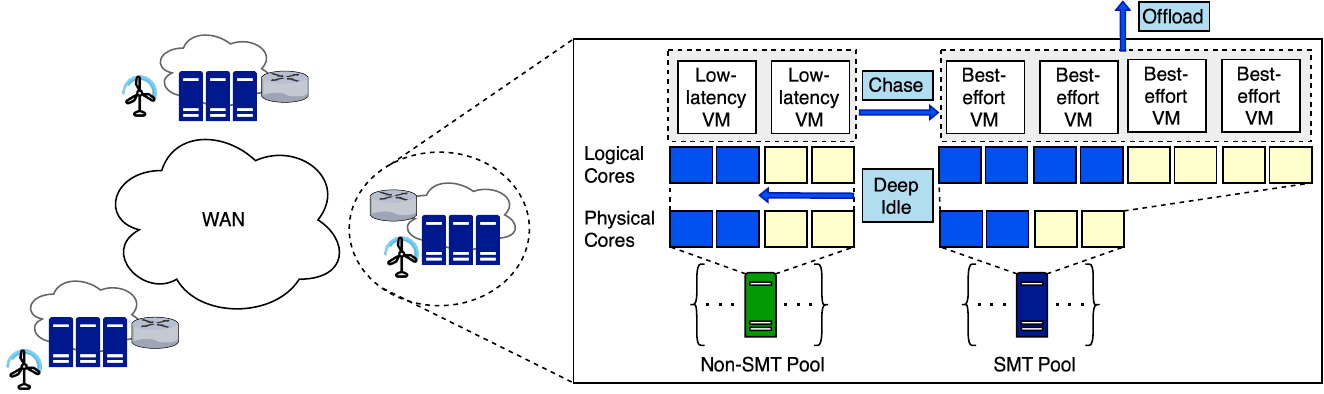}
    \caption{System architecture of the proposed load shifting technique.}
    \label{fig::system-model}
\end{figure*}

Figure \ref{fig::system-model} illustrates our system model. We consider a multi-region cloud deployment. Each region integrates intermittent renewable energy in the data center power delivery. As a result, data center servers in each cloud region undergo dynamic energy capacity availability. In return, server compute capacity dynamically scales up and down. We model that with the dynamic server resource scaling model. Workload schedulers in cloud regions match their workload execution with the server resource availability via load shifting both within the data center and data centers across regions via VM offloading over a Wide Area Network (WAN). Regions receive workloads as VM arrivals, which are either VMs of low-latency applications or VMs of best-effort applications. The WAN that connects regions offer network routes with varying latency performance, which we model with the latency performance model. Using the derived models, we then formulate our problem.

\par\noindent\textbf{Server dynamic resource scaling model:} We model server resource scaling from the dynamic availability of renewable energy capacity through availability of CPU cores. Given the energy capacity, server power is managed by enabling a sufficient amount of CPU cores. The same technique has been commonly used as a core-level server power management mechanism \cite{li2011solar-core,agarwal2023slackshed}.

A data center in our model integrates a mixed power delivery of both carbon-intensive stable power sources, such as fossil fuel-based energy generation, and low-carbon-intensive intermittent renewable energy sources, such as solar and wind. Stable sources provide baseline energy capacity for the data center, whereas intermittent sources provide peaks and valleys of low-carbon-intensive energy availability. Data center power consumption is typically a combination of information and technology components (IT) such as servers, cooling, and internal power conditioning systems \cite{ahmed21dc-energy-review}. Servers in our system model facilitate CPU-intensive low-latency and best-effort workloads. Therefore, majority of dynamic power draw variations corresponds to CPU power. For such systems, prior works show server power can be estimates as a linear function of CPU power with 90\% accuracy \cite{radovanovic2022dcpowermodel}. Moreover, for multi-core CPUs, the cumulative sum of core power becomes a close upper bound to the CPU power \cite{basmadjian2012multicorepowermodel}. Combining both, we use the following model to estimate server power.

\begin{equation}
\label{eq:core-power}
    P_{S}(t) = f(\sum_{i=1}^{N} P_{CORE_i}(t))
\end{equation}
where at time $t=t$, $P_{S}(t)$ is the server power, $N$ is the number of homogeneous cores, $P_{CORE_i}(t)$ is the power consumption of the $i^{th}$ core, and $f$ is a linear function.

We set server power capacity of the stable grid energy sources ($P_{grid}$), such that grid capacity is sufficient to support both non-IT power load and half of the peak IT power load.

\begin{equation*}
P_{grid} = P_{S_{peak}} / 2
\end{equation*}

We set power capacity from the renewable sources ($P_{rnw}(t)$) as a two level power signal, where once sufficient capacity is available, renewable sources are able to provide power to support peak server load.

\begin{equation*}
P_{rnw}(t) = \begin{cases}
    P_{S_{peak}} / 2, & \text{t = High renewables} \\
    0, & \text{t = Low renewables}
\end{cases} 
\end{equation*}

With that, we define our dynamic server capacity scaling model ($P_{cap}(t)$) as,

\begin{equation*}
P_{cap}(t) = P_{grid} + P_{rnw}(t)
\end{equation*}

\par\noindent\textbf{WAN latency performance model: }We model WAN latency performance based on the traffic management characteristics of WANs connecting multi-region cloud deployments \cite{miao24megate}. In that, communication latency across WAN is unpredictable, thus changes overtime. With that, we model WAN latency performance as follows.

For the duration of $\Delta T$, the set of executed low-latency VMs ($S_{VM}$) is,

\[
S_{VM} = \{VM_1,...,VM_m\}
\]

For $S_{VM}$, we calculate WAN latency performance ($Lat(S_{VM})$) as,

\[
Lat(S_{VM}) = \frac{1}{\sum_{i}^{m} \int_{t}^{\Delta T} L_i(t) \,dt} 
\]
where $L_i(t)$ is the end-user latency of the $i^{th}$ VM at time $t$.

\par\noindent\textbf{Problem formulation:} We formulate our problem as follows. Given an arbitrary time period $\Delta T$, maximize end-user latency performance of VMs while meeting the dynamic capacity scaling of servers.

\[
\text{Maximize} \quad Lat(S_{VM}) \quad \text{and} \quad  \forall S \in \text{Servers}, P_S(t) \leq P_{cap}(t)
\]

\section{Load Shifting for Low-latency Applications with SMT Core Pooling}
\label{sec::technique}

In this section, we detail the design of our load shifting technique. Our technique addresses the research questions identified in Section \ref{sec::background-and-motivation} to accommodate low-latency applications in cloud load shifting. In Section \ref{sec::performance-evaluation} we showcase its superiority in delivering improved latency performance for low-latency applications.

Figure \ref{fig::system-model} illustrates the system architecture of the hardware-software co-design of the proposed technique. At the hardware level, we leverage heterogeneous server pools of SMT cores to maintain a static CPU capacity amidst renewable supply dynamics. We employ core-level power management to meet the dynamic resource scaling of servers for renewable supply variations. In return, we maintain a static set of logical cores cores across the server pools. We then exploit the static set at the software layer. Using a novel VM scheduling algorithm, we efficiently manage low-latency application VMs for the static set, lowering their possibilities of shifting over the WAN. Our approach involve three steps: 1) Core-level power management to meet server resource scaling, 2) SMT pooling to yield static resource availability, and 3) novel VM scheduling algorithm to utilize SMT pooling for low-latency VMs. Collectively, we maximize retaining low-latency VMs within the local region by exploiting the hardware multi-threading of SMT.

\subsection{Hardware-level logical core management}

\noindent\textbf{Server power management:} To cater to dynamic server resource scaling for the supply dynamics of renewable energy, we employ per-core deep idling \cite{rafael2018cpu-idle} to manage the server power draw. At $t=\text{Low Renewables}$, we set half of the server cores into the deep sleep, such that the peak power draw of the server meets $P_{cap}(t)$.

\[
P_S(t=\text{Low Renewables}) \leq f(\sum_{i=1}^{N/2} P_{CORE_{peak}}) 
\]
where $P_{CORE_{peak}}$ is the peak power draw of each homogeneous core. Opposed to that, we awake all cores to better utilize $P_{cap}(t)$ at $t=\text{High Renewables}$.

\[
P_S(t=\text{High Renewables}) \leq f(\sum_{i=1}^{N} P_{CORE_{peak}}) 
\]

We then leverage CPU Simultaneous Multi-threading (SMT) to exploit our server power management to yield a static set of CPU resources. 

\noindent\textbf{SMT pooling:} Simultaneous Multi-threading (SMT) in CPUs doubles the available number of logical cores \cite{openstacktopologies} while providing better performance due to its hardware multi-threading nature. To exploit our server power management with SMT, we maintain two server pools where the servers in one of the pools enable SMT (i.e., SMT pool). The number of logical cores in servers of the SMT pool is double the amount of its physical CPU cores. In combination with our server power management, which deep idles half of the physical cores at renewable supply valleys, we realize a static set of logical cores across the server pools, regardless of the renewable supply state. At supply valleys, servers in the SMT pool provide the static set of logical cores, whereas, at supply peaks, servers in the non-SMT pool provide that instead. An example of that is illustrated in Figure \ref{fig::system-model}. In that example, $N$ is set to $4$. At supply peaks, the non-SMT pool exhibits four logical cores, and the SMT pool exhibits eight logical cores. At supply valleys, the SMT pool exhibits four logical cores. Collectively, a static set of four logical cores is present in the local region at all times, which is a significant advantage over traditional server power management, where the renewable energy integration often down scale resource capacity of each server.

The primary reason for not enabling SMT in both pools is that core oversubscription methods such as SMT can still impact the low-latency performance of VMs \cite{openstack2023realtime}. In our system architecture, the static set of logical cores only rely on SMT cores during energy supply valleys, thus the latency performance impact is further reduced. In the next subsection, we efficiently utilize the static set of logical cores for low-latency applications via a novel VM scheduling algorithm. 

\subsection{Software-level VM scheduling algorithm}

\begin{algorithm}
\caption{Proposed VM Scheduling Algorithm for SMT Pooling.}
\label{alg:vm-packing}
\begin{algorithmic}[1]
    \renewcommand{\algorithmicrequire}{\textbf{Input:}}
    \renewcommand{\algorithmicensure}{\textbf{Output:}}

    \REQUIRE 
        \begin{itemize}
            \item $v$: Incoming VM request.
            \item $S_{\text{smt}}$: Set of servers in the SMT pool.
            \item $S_{\text{non-smt}}$: Set of servers in the non-SMT pool.
            \item $P(t) \in \{\mathit{peak}, \mathit{valley}\}$: Renewable supply state at time $t$.
            \item $P(t^-) \in \{\mathit{peak}, \mathit{valley}\}$: Renewable supply state before time $t$.
        \end{itemize}

    \ENSURE Scheduling decisions for incoming VM placement and server pool management.

    \STATE \textbf{Function} \textsc{HandleVMPlacement}($v$, $P(t)$): \label{algo::func::handle-vm-placement}
    \STATE \quad \textbf{if} \textsc{IsLowLatency($v$)} \textbf{then} \label{algo::func::is-low-latency}
    \STATE \quad \quad \textbf{if} $P(t) = \mathit{peak}$ \textbf{then}
    \STATE \quad \quad \quad \textsc{Place($v,S_{\text{non-smt}}$)}
            \quad 
    \STATE \quad \quad \textbf{else if} $P(t) = \mathit{valley}$ \textbf{then}
    \STATE \quad \quad \quad \textsc{NotAdmit($v$)} \label{algo::func::not-admit}
            \quad 
    \STATE \quad \textbf{end if}
    \STATE \quad \textbf{else} \quad \label{algo::func::be-vm-placement}
    \STATE \quad \quad \textbf{if} $P(t) = \mathit{peak}$ \textbf{then}
    \STATE \quad \quad \quad \textsc{Place($v,S_{\text{smt}}$)} \label{algo::func::limit-be-to-smt-at-peak}
            \quad 
    \STATE \quad \quad \textbf{else if} $P(t) = \mathit{valley}$ \textbf{then}
    \STATE \quad \quad \quad \textsc{Place($v,S_{\text{smt}} \cup S_{\text{non-smt}}$)}
            \quad 
    \STATE \quad \textbf{end if}
    \STATE \textbf{end function}

    \STATE \textbf{Function} \textsc{HandlePowerTransition}($P(t^-)$, $P(t)$): \label{algo::func::handle-power-transition}
    \STATE \quad \textbf{if} $P(t^-) = \mathit{peak}$ \textbf{and} $P(t) = \mathit{valley}$ \textbf{then}
    \STATE \quad \quad \textsc{OffloadBestEffortVMs($S_{\text{smt}}$)} \label{algo::func::clean-smt}
            \quad 
    \STATE \quad \quad \textsc{LiveMigrate($v\in\text{Low-latency}$, $S_{\text{non-smt}}$, $S_{\text{smt}})$}
            \quad 
    \STATE \quad \textbf{else if} $P(t^-) = \mathit{valley}$ \textbf{and} $P(t) = \mathit{peak}$ \textbf{then} \label{algo::func::clean-non-smt}
    \STATE \quad \quad \textsc{OffloadBestEffortVMs($S_{\text{non-smt}}$)}
            \quad 
    \STATE \quad \quad \textsc{LiveMigrate($v\in\text{Low-latency}$, $S_{\text{smt}}$, $S_{\text{non-smt}}$)} 
            \quad
    \STATE \quad \textbf{end if}
    \STATE \textbf{end function}

    \STATE \textbf{Main Procedure}:
    \STATE \quad \textbf{while} \textsc{SystemIsRunning()} \textbf{do}
    \STATE \quad \quad \textbf{if} \textsc{IsPowerTransition}($t^-, t$) \textbf{then}
    \STATE \quad \quad \quad \textsc{HandlePowerTransition}($P(t^-), P(t)$)
    \STATE \quad \quad \textbf{end if}
    \STATE \quad \quad \textbf{for each} $v$ \textbf{do}
    \STATE \quad \quad \quad \textsc{HandleVMPlacement}($v, P(t)$)
    \STATE \quad \quad \textbf{end for}
    \STATE \quad \textbf{end while}

\end{algorithmic}
\end{algorithm}

In this section, we propose a software-level VM scheduling algorithm to utilize the hardware-level logical cores provided by SMT pooling. Algorithm \ref{alg:vm-packing} outlines our VM scheduling algorithm. It takes incoming VM placement requests and the state of the renewable energy supply (i.e., peak or valley) as inputs. It then determines VM placement decisions and handles server pool management for renewable energy dynamics.

The pseudo-code for VM placement is outlined in the subroutine \textsc{HandleVMPlacement} (line \ref{algo::func::handle-vm-placement}). It first identifies the criticality of the VM as either low-latency or best-effort (line \ref{algo::func::is-low-latency}). For low-latency VM placement requests at peak renewable energy supply, requests are admitted to the cloud region for placement and deployed in the non-SMT pool. Since CPU cores of the non-SMT pool do not have the performance overhead of having hardware multi-threading, the placement decision aims for maximum CPU performance. Conversely, requests will not be admitted to the cloud region if the renewable energy supply is at a valley (line \ref{algo::func::not-admit}). At that stage, available cores in the non-SMT pool are reserved to restore the performance of already deployed low-latency VMs at renewable energy peaks. We provide inner details of the server pool management in the next paragraph. In the case of best-effort VM placement requests (line \ref{algo::func::be-vm-placement}), we admit and deploy those regardless of the renewable energy supply state, with the exception of limiting their deployment to the SMT pool during supply peaks (line \ref{algo::func::limit-be-to-smt-at-peak}).

The pseudo-code for server pool management is outlined in the subroutine \textsc{HandlePowerTransition} (line \ref{algo::func::handle-power-transition}). It takes two input parameters: the state of the renewable energy supply at times $t$ and immediately before ($t^-$). In case of a transition from supply peak to a valley, we first offload best-effort VMs in the SMT pool from the cloud region (line \ref{algo::func::clean-smt}) and internally live migrate low-latency VMs to the SMT pool from the non-SMT pool. Here, offloaded VMs are handled similarly to standard load shifting across cloud regions, which places those VMs outside the local cloud region. Our approach exploits the best-effort nature of VMs to relax constraints in offloading. In the case of a transition from supply valley to a peak, we first offload best-effort VMs from the non-SMT pool and live migrate VMs from the SMT pool to the non-SMT pool (line \ref{algo::func::clean-non-smt}). Overall, through live migrations, server pool management maintains deployed low-latency VMs inside the data center. Migrations cause minimum disruption to the application execution, and most importantly, end-user latency performance is kept intact by keeping VMs in the same cloud region throughout. We make room for the low-latency VMs via offloading best-effort VMs, aiming for a minimum impact from geographical load shifting towards the application service quality.

Our algorithm's collective management of VM placement and server pool management is designed to improve the service quality of the cloud region's already admitted low-latency VMs. It improves the utilization of SMT pooling and the server power management's approach of halving the number of available physical cores, guaranteeing a fixed resource capacity for low-latency VMs through their live migrations and offloading best-effort VMs.

\section{Implementation}
\label{sec::implementation}

We implement our proposed technique in a real experimental cloud environment. We implement core-level power management of servers via the CPU idle states feature \cite{intel18idle-time-mgt} and SMT pooling using the same feature present in server CPUs \cite{intel25ht}, both for Intel CPUs. We implement our proposed VM scheduling algorithm at the cloud resource management layer using OpenStack \cite{openstack-scheduling}.

First, we implement core-level power management of the server by deploying a daemon service in each server \cite{tharindu25core-power-mgt}. The daemon service wraps the Intel power optimization library that provides low-level API control of the CPU c-states in each CPU core. We then expose high-level RESTful APIs to the deep idle half of the CPU cores. As a result, upon receiving a core deep idle request, the daemon service overrides the default kernel's behavior of CPU idle states and maintains half of the CPU cores at the deepest c-state. Conversely, a wake request will restore the deep idle state. We write the daemon service in golang. Secondly, we select the set of servers for the SMT pool and enable CPU hyper-threading through their BIOS settings. Finally, we deploy OpenStack to manage the servers. We modify its default server filter in the VM scheduling workflow to omit specific server pools for certain VM types (see Algorithm \ref{alg:vm-packing}). We assume VM placement requests that are unsuccessful in finding a scheduling decision will be offloaded from the cloud region. Further, we implement a Golang controller in the OpenStack control plane for server pool management. It is driven by the events from the renewable energy supply, for which we expose a high-level API. Upon triggering, it conducts live migrations between servers and emulates VM offloading by conducting VM evictions through OpenStack APIs (see Algorithm \ref{alg:vm-packing}).

\section{Performance Evaluation}
\label{sec::performance-evaluation}

In this section, we evaluate the performance of our proposed technique. We outline our experimental design and setup, compare our results with the state-of-the-art baseline, and analyze them in detail.

\subsection{Experimental design and setup}

We conduct evaluation experiments in a real prototype multi-node cloud region. We model the multi-region characteristics for our experiments using real measured data from production cloud regions \cite{miao24megate}. In our prototype cloud, we allocate an identical HP ProLiant server to each SMT and non-SMT pool. Each server has an Intel Xeon CPU with 12 physical cores. For the SMT pool, we enable Intel Hyper-threading through its bios settings. Both servers are part of a research server cluster in a private network and share a fast network fabric. We install OpenStack in both servers, marking one as the control plane. We deploy our Golang daemon service on both servers and the Golang controller on the server marked for OpenStack's control plane.

\subsubsection{Baselines:} We compare our proposed technique with state-of-the-art \textbf{\textit{Space-Shifting}}. Space-shifting is the predominantly used load shifting technique \cite{radovanovic23datacentercarbonaware, murillo24cdn-shifter}. It aims to manage the cloud region's power draw by shifting its flexible workloads across space to other cloud regions based on energy availability. We use space-shifting to compare the superiority of our proposed technique in accommodating low-latency workloads.

\subsubsection{Workloads:} We use Microsoft Azure's VM packing data \cite{azure2020packingtrace} to expose our testbed to realistic VM requests in production clouds. We scale Azure's data to match our experimental deployment by sampling and synthesizing VM arrival traces using the data. For renewable energy supply dynamics, we use solar dynamics data from the ELIA dataset \cite{eliaopendata}. Further, we use real latency variation data measured for cloud regions \cite{miao24megate} in our experimental setup to evaluate VM latency impact from shifting.

\subsubsection{Metrics:} To measure the latency performance, we use two latency metrics. We use \textbf{\textit{application latency performance}} to measure the performance impact inside the low-latency VM. We use \textbf{\textit{end-user latency}} to measure the impact of shifting VMs across the WAN. To measure application latency performance, rather than measuring the latency performance of specific cloud applications, we monitor the latency performance of the VM's guest operating system. For that, we use the cyclictest tool \cite{cyclictest}. For end-user latency, we use WAN latency data. For that, we measure latency for the duration of the VM's lifetime provided by the trace data. Using a statistical model we build from WAN latency data, we sample a latency value for each time step and get the aggregated value. To measure the scheduling performance, we monitor the number of \textbf{\textit{Retained}}, \textbf{\textit{Offloaded}}, and \textbf{\textit{Not Admitted}} VMs in the cloud region. Retained VMs complete their lifetime inside the cloud region, offloaded VMs are interrupted at mid-life to get shifted over WAN, and not-admitted VMs are diverted to a different cloud region at admittance.

\begin{figure}[tpb]
    \centering
    \includegraphics[width=0.9\linewidth]{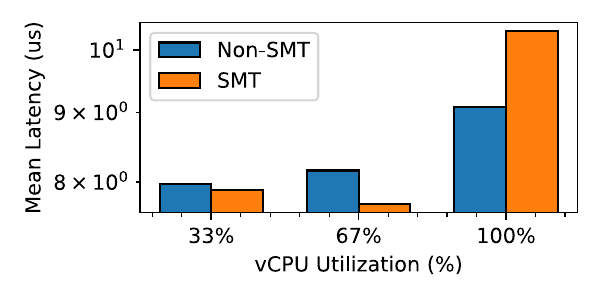}
    \caption{Comparison of mean application latency performance of heterogeneous server pools as resource contention increases.}
    \label{fig::rslt::lat-vs-resource-contention}
\end{figure}

\subsection{Results and Analysis}

We evaluate the performance of our proposed technique using a two-fold approach. Firstly, we evaluate the performance impact of executing VMs on the heterogeneous server pools. Secondly, we evaluate the performance impact of various placement decisions made due to the load-shifting approach.

\noindent\textbf{Application latency performance with SMT pooling}: Due to SMT pooling, VMs in our technique can be deployed on either SMT or non-SMT cores. To evaluate application latency performance, we measure the latency performance of the guest operating system for various core allocation configurations. Since SMT cores share CPU components, our experiment is designed to measure the performance impact as resource contention in the CPU increases. We first allocate VMs to occupy a portion of the CPU cores available in a server of each pool. Then, we execute a system load in the VM to emulate a utilized application and measure the latency performance of the guest operating system. We then repeat the experiment for different allocated CPU core portions until it reaches 100\%. Our experiment measures the low-latency application impact of deploying a VM in a server for different packing levels across the server pools. Figure \ref{fig::rslt::lat-vs-resource-contention} illustrates the results. It shows the mean latency value measured for the guest operating system over the utilization of the server's virtual CPU cores (vCPU) across both SMT and non-SMT pools.

The results show that the impact of application latency from deploying a VM on a specific server pool significantly depends on CPU utilization. For both 33\% and 67\% of the vCPU utilization, the mean latency performance of VMs on SMT and Non-SMT pools is around 8 microseconds. However, as utilization increases to 100\%, there is a significant increase in the VM mean latency, where packing on SMT cores increases from 8 to 10 microseconds. Packing on non-SMT cores increases from 8 to only about 9 microseconds. In comparison, packing on SMT cores increases the mean latency by 11.97\%. In this context, the application latency impact of placing a VM across heterogeneous server pools depends on the service quality of the application. For instance, server pooling will primarily impact an application sensitive to 1-microsecond performance degradation. In contrast, low-latency applications that can tolerate larger latency penalties will not exhibit degradation in their service quality.

\begin{figure*}[t]
    \centering
    \subfloat[Comparison for Best-effort VMs%
              \label{fig::rslt::summary::best-effort}]%
             {\includegraphics[width=0.45\linewidth]{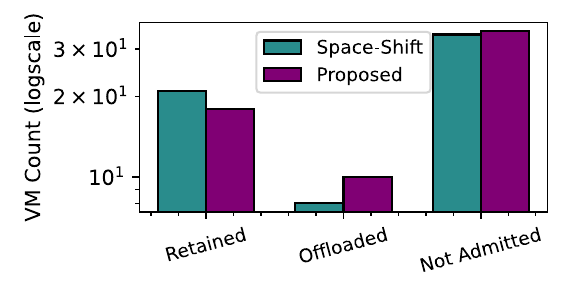}}\hfill
    \subfloat[Comparison for Low-latency VMs%
               \label{fig::rslt::summary::low-latency}]%
             {\includegraphics[width=0.45\linewidth]{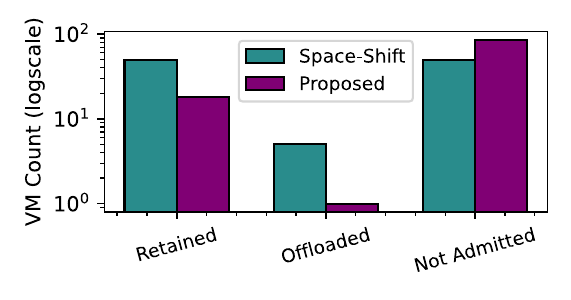}}
    \caption{Comparison of VM-scheduling performance in the experimental cloud region.}
    \label{fig:rslt:summary}
\end{figure*}

\begin{figure}[tpb]
    \centering
    \includegraphics[width=0.7\linewidth]{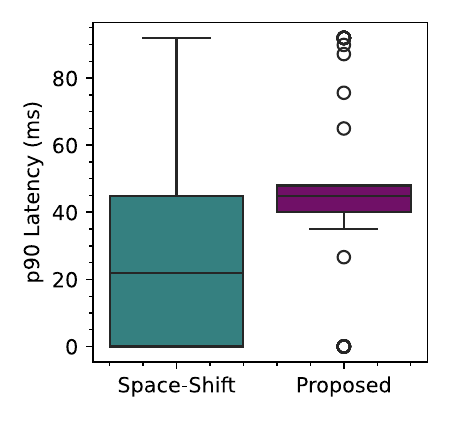}
    \caption{Comparison of p90 end-user latency distribution of Low-latency VMs.}
    \label{fig::rslt::p90-lat-dst}
\end{figure}

\noindent\textbf{Scheduling impact of low-latency VMs}: In this experiment, we measure the scheduling impact of retaining deployed VMs inside the cloud region during its lifetime, offloading VMs in mid-life to other cloud regions, and VM admittance to the cloud region. We replay VM request arrivals of the Azure trace and 24-hour renewable supply change dynamics. We define a threshold in the renewable energy supply to define peaks and valleys. Afterward, we conduct the same experiment for our proposed technique and the baseline. Figure \ref{fig::rslt::summary::best-effort} and \ref{fig::rslt::summary::low-latency} illustrates the results for each best-effort and low-latency VM types. It shows the number of VMs in each category of the x-axis.

The results show that the proposed technique significantly surpasses Space-Shift in reduced offloaded events for low-latency VMs with an 80\% reduction. In comparison, it increases the best-effort VM offloading. In admitting VM requests for deployment, the proposed technique is closely the same, with a slightly decreasing number of not-admitting events for both best-effort and low-latency VM types. The Space-Shift approach performs better in retaining both low-latency and best-effort VM types. Collectively, the behavior of the proposed technique shows its superiority in managing low-latency VMs. This is because prior studies show that even with relatively degraded performance, users favor a deterministic application performance \cite{kumhare2021poweroversubscription}. In that regard, offloading events incur the most disruption to low-latency applications since offloading can introduce latency overheads from the wide area network's performance. The proposed technique shows the minimum offloading events and compromises in not admitting VMs. Although not admitting does shift the VM across WAN, end-user service quality would indicate the degraded performance since deployment rather than showing mid-life.

\noindent\textbf{End-user latency impact of low-latency VMs}: We measure the impact of end-user latency in our scheduling impact experiments to evaluate the impact of shifting VMs across WAN. Using probabilistic models, we generate data from real inter-cloud-region latency over WAN. We sample a distribution of latency values for each VM for the lifetime and calculate its p90 value. WAN latency does not impact retained VMs, which complete their entire lifetime inside the local cloud region. However, WAN latency does impact the not-admitted VMs and has a partial impact on offloaded VMs. Figure \ref{fig::rslt::p90-lat-dst}  illustrates the results. It compares the distributions of p90 latencies of VMs.

The results show the superiority of the proposed technique in reducing the p90 latency variance. Although there are several outliers, the latency performance of VMs with the proposed technique primarily concentrates around 40 milliseconds with a coefficient of variation of 0.62. In contrast, with Space-Shift, the p90 latency value disperses around 20 milliseconds with a coefficient of variation of 1.10. The proposed technique reduces the coefficient of variation of p90 latency by 43.81\%. Therefore, the results indicate that VMs managed with the Space-Shift technique will most likely deliver unpredictable end-user performance compared to the proposed approach, which is unfavorable for the service quality \cite{kumhare2021poweroversubscription}.

\section{Related Work}
\label{sec::related-works}

\begin{table*}[t]
  \centering
  \caption{Comparison of relevant works with our proposed technique.}
  \label{tab:comparison}
  \renewcommand{\arraystretch}{1.1} %
  \setlength{\tabcolsep}{5pt}      %
  \begin{tabular}{%
    p{0.18\linewidth}|
    >{\centering\arraybackslash}p{0.14\linewidth}|
    >{\centering\arraybackslash}p{0.14\linewidth}|
    >{\centering\arraybackslash}p{0.14\linewidth}|
    >{\centering\arraybackslash}p{0.14\linewidth}|
    >{\centering\arraybackslash}p{0.12\linewidth}
  }
    \hline
    \textbf{Work} 
    & \textbf{Multi-cloud Carbon Optimization} 
    & \textbf{Load Shifting}
    & \textbf{Uninterruptible Execution} 
    & \textbf{Low-latency Applications} 
    & \textbf{Static Resource Capacity}
    \\ 
    \hline
     Radovanovic'23 \cite{radovanovic23datacentercarbonaware} 
      & \checkmark
      & \checkmark
      &
      &
      &
      \\ 
    Carbonscaler'23 \cite{hanafy23carbonscaler} 
      & \checkmark
      & \checkmark
      & \checkmark
      &
      &
      \\ 
    Zheng'20 \cite{zheng2020mitigating} 
      & \checkmark
      & \checkmark
      & \checkmark
      &
      &
      \\ 
     CDN-Shifter'24 \cite{murillo24cdn-shifter} 
      & \checkmark
      & \checkmark
      & \checkmark
      & \checkmark
      &
      \\ 
    \textbf{Our Proposed} 
      & \checkmark 
      & \checkmark 
      & \checkmark 
      & \checkmark
      & \checkmark
      \\
    \hline
  \end{tabular}
\end{table*}

\textbf{Integrating renewables in clouds:} Many previous studies have explored integrating renewable energy into cloud data center operation. They commonly exploit load shifting techniques in both space and time, such as geographical load balancing \cite {radovanovic23datacentercarbonaware,murillo24cdn-shifter,hanafy23carbonscaler}, VM migrations \cite{masanet2020recalibratingdatacenterenergyestimates, radovanovic23datacentercarbonaware, liu2011greeningloadbalance, zheng2020mitigating, james2019lowcarbonkubenetes}, and workload admissions and capacity planing \cite{radovanovic23datacentercarbonaware}. Most techniques are intended for flexible workloads that can tolerate delayed execution and responses. For instance, carbon optimization at Google \cite{radovanovic23datacentercarbonaware} leverage internal workloads that can withstand delays as long as 24 hours, such as machine learning, data processing, and simulation. They employ virtual capacity curves across their geographically dispersed data centers and employ temporal shifting to perform carbon-optimized computing. Some works explore shifting workloads across locations over WAN, considering the impact of WAN's dynamic traffic congestion towards the workloads \cite{guo2023carbon}. However, they do not focus on low-latency applications and only focus migration cost calculations for the energy cost of the transfer. In contrast, we explore geographical load shifting to accommodate low-latency applications and provide a technique to better manage server power over intermittent renewable energy while maintaining service quality of low latency applications.
\\ 
\textbf{VM scheduling of dynamic inventories:} Cloud VM scheduling for energy management is a widely researched problem \cite{hadary2020protean, baker2018cloud}. Many aim to optimize VM scheduling to increase energy efficiency for static inventories, where server resources remain static and server power is managed through the workload management \cite{liu2022dynabin}. However, recent works explore VM scheduling for dynamic inventories where available server resources are adjusted to tolerate the changes in the data center power delivery. For instance, Microsoft increase utilization of their clouds via power oversubscription, where server are provisioned to oversubscribe the power capacity \cite{kumhare2021poweroversubscription} and power overdraw events are managed by throttling the server compute capabilities. They leverage a dynamic VM scheduling technique to reduce the workload impact, considering VM priority levels. However, workload throttling in their approach can lead to non-deterministic application performances. In our work, we employ a dynamic inventory to absorb renewable intermittencies and conduct VM scheduling to leverage the idea of chasing static resources within the datacenter for low-latency applications, providing better stability in application performance.

\section{Conclusions and Future Work}
\label{sec::conclusion-future-work}

Although load shifting is commonly used in carbon-optimized clouds to utilize intermittent renewable energy availability across regions, it often overlooks low-latency applications due to dynamic latency overheads of wide area networks (WAN) connecting regions. To this end, our work proposes a load-shifting technique to accommodate low-latency applications with minimum impact. The proposed technique employs heterogeneous server pools differentiated by simultaneous multi-threading (SMT) feature. It exploits SMT pools with core-level power management to maintain a static set of logical cores across server pools, which then is utilized with a VM scheduling algorithm to prevent offloading low-latency VMs from the local cloud regions, avoiding costly communication overheads of WAN. We implement the proposed technique with OpenStack and CPU idle states and evaluate its performance using Azure VM arrival traces on an experimental cloud region with an HP ProLiant server allocated to each SMT and non-SMT pool. In comparison to state-of-the-art space-shifting, our empirical results demonstrate the superiority of the proposed technique with an 80\% reduction in offloading low-latency VMs, 43.81\% reduction in coefficient of variation of end-user latency, and worst-case latency compromise of 11.97\% due to SMT cores. 

In future work, we plan to advance the VM scheduling algorithm in the proposed technique to utilize fine-grain allocations of non-SMT cores during renewable energy valleys. Further, tuning the proposed technique for application-specific requirements can enable relaxed constraints of VM scheduling, which can further present opportunities to improve CPU utilization.

\bibliographystyle{IEEEtran}
\bibliography{ref}

\vfill

\end{document}